\documentclass[10pt,superscriptaddress,twocolumn,amsmath,amssymb,aps,prl,showpacs]{revtex4}
\usepackage{graphicx}
\usepackage{dcolumn}
\usepackage{bm}
\usepackage{amsmath}
\usepackage{paralist}
\usepackage{amsmath,amssymb,mathrsfs}
\usepackage{braket}
\usepackage{hyperref}
\usepackage{pdfpages}

\newcommand{\Li}{\mathrm{Li}}
\def\be{\begin{equation}}
\def\ee{\end{equation}}
\def\bea{\begin{eqnarray}}
\def\eea{\end{eqnarray}}

\begin{document}

\title{Universal Properties of Fermi Gases in One-dimension\footnote{The authors WBH and YYC contributed equally to the calculations in this paper.}}

\author{Wen-Bin He}
\affiliation{State Key Laboratory of Magnetic Resonance and Atomic and Molecular Physics,
Wuhan Institute of Physics and Mathematics, Chinese Academy of Sciences, Wuhan 430071, China}
\affiliation{University of Chinese Academy of Sciences, Beijing 100049, China.}

\author{Yang-Yang Chen}
\affiliation{State Key Laboratory of Magnetic Resonance and Atomic and Molecular Physics,
Wuhan Institute of Physics and Mathematics, Chinese Academy of Sciences, Wuhan 430071, China}
\affiliation{University of Chinese Academy of Sciences, Beijing 100049, China.}

\author{Shizhong Zhang}
\email[]{shizhong@hku.hk}
\affiliation{Department of Physics and Centre of Theoretical and
Computational Physics, The University of Hong Kong, Hong Kong, China}

\author{Xi-Wen Guan}
\email[]{xiwen.guan@anu.edu.au}
\affiliation{State Key Laboratory of Magnetic Resonance and Atomic and Molecular Physics,
Wuhan Institute of Physics and Mathematics, Chinese Academy of Sciences, Wuhan 430071, China}
\affiliation{Center for Cold Atom Physics, Chinese Academy of Sciences, Wuhan 430071, China}
\affiliation{Department of Theoretical Physics, Research School of Physics and Engineering,
Australian National University, Canberra ACT 0200, Australia}

\date{\today}



\begin{abstract}
In this Rapid Communication, we investigate the universal properties of a spin-polarized two-component Fermi gas in one dimension (1D) using Bethe ansatz. We discuss the quantum phases and phase transitions by obtaining exact results for the equation of state, the contact, the magnetic susceptibility and  the contact susceptibility, giving a precise understanding of the 1D analogue of the Bose-Einstein condensation and Bardeen-Cooper-Schrieffer crossover in three dimension (3D) and the associated universal magnetic properties. In particular, we obtain the exact form of the magnetic susceptibility $\chi \sim  {1}/{\sqrt{T}}\exp(-\Delta/T)$ at low temperatures, where $\Delta$ is the energy gap and $T$ is the temperature. Moreover,  we establish exact upper and lower bounds for the relation between polarization $P$ and the contact $C$ for both repulsive and attractive Fermi gases. Our findings emphasize the role of the pair fluctuations in strongly interacting 1D fermion systems that can shed light on higher dimensions.
\end{abstract}
\maketitle

In the past decade, remarkable progresses have been made in the study of strongly interacting fermions in three-dimensional (3D) Bose-Einstein condensate(BEC)-Bardeen-Cooper-Schrieffer (BCS) crossover~\cite{Zwerger:2012,Chin:2004,Zwierlein:2005,Zwierlein:2006}, including precise measurements of equations of states (EOS)~\cite{Nascimbene:2010,Navon:2010,Horikoshi:2010,Ku2012},  the investigations of magnetic and pairing/depairing phenomena~\cite{Partridge:2006,Shin:2006,Olsen:2015} and polaron \cite{Polaron-1,Polaron-2,Polaron-3,Kohstall:2012}  in a spin polarized system. In addition, the 2D Fermi gas has also been realized and attracted a lot of attention both experimentally~\cite{Polaron-4,Frohlich:2009,Makhalov:2014,Ong:2015,Ries:2015,Muthy:2015,Boettcher:2015,Fenech:2015} and theoretically~\cite{Liu:2013,Anderson:2015,Bauer:2014,Mulkerin:2015,Ngampruetikorn:2013}. Despite these advances, fundamental questions still remain. In particular, what is the role of pair fluctuations on thermodynamics and transport phenomena, especially in the vicinity of transition temperature at unitarity.

Similar questions can be answered in a more affirmative manner in the case of 1D Fermi gases due to the existence of exactly solvable models. Recently, dramatical progresses have been made in the experimental realizations of many exactly solvable models of 1D interacting bosons and fermions~\cite{Cazalilla,GuaBL13}. One naturally expects to gain deeper insight into the 1D analogue of the 3D BEC-BCS crossover as well as  the associated universal thermodynamics from the Bethe ansatz perspective. In fact, the precise equation of state obtained from the  thermodynamic Bethe ansatz (TBA) equations~\cite{Tak70} not only gives rise to important characteristics of the Tomonaga-Luttinger liquids (TLL) but also provides universal laws,  such as quantum scalings, dimensionless ratios, Tan's contact and universal relations between macroscopic properties~\cite{Tan:2008,Guan:2013PRL,YCLRG:2015}, which could shed new light on many-body phenomena in higher dimensions. However, a complete derivation of the  thermodynamics of an arbitrarily polarized gas and the role of pair fluctuation in 1D systems still remain a  challenging problem in the field of cold atoms~\cite{GuaBL13,Anderson:2015,Loheac:2015,Patu:2015}.

In this Rapid Communication,  we show that the TBA equations of the 1D spin-$1/2$ Fermi gases can serve as a  framework to derive benchmark thermodynamics  for a wide range of physical phenomena. 
Analytical and numerical results are obtained for the above mentioned  key physical quantities that cover a wide range of interaction and temperature regimes. In particular, we obtain exact results for the magnetic susceptibility and prove its universal behavior at low temperature $\chi \sim  {1}/{\sqrt{T}}\exp(-\Delta/T)$, where $\Delta$ is the energy gap of the system and $T$ is the temperature. To characterize the strength of pair fluctuations, we define the ``contact susceptibility" and show that it captures the crossover from the TLL regime to the quantum critical regime. We further establish the upper and lower bounds for the contact in terms of the polarization $P$ of the system, and show how it ranges from the random collision limit (high temperature limit) $C \sim 1-P^2$ to the fully paired state $C \sim 1-P$ at low temperatures~\cite{Bardon:2014}.

{\em The model.} A two-component Fermi gas with contact interaction in one-dimension~\cite{Yang,Gaudin} (the Yang-Gaudin model) is described by the Hamiltonian $\mathcal{H} = \mathcal{H}_0- \mu N - HM$, where $\mathcal{H}_0 =-\frac{\hbar^{2}}{2m}\sum_{i=1}^{N}\frac{\partial^{2}}{\partial x_{i}^{2}}+g_{\rm 1D}\sum_{1\leq i<j\leq N}\delta(x_{i}-x_{j})$. $N$ is the total number of particles and $M=(N_{\uparrow}-N_{\downarrow})/2$ is the spin polarization. $\mu$ is the chemical potential and $H$ is the magnetic field  (chemical potential difference). The effective 1D interaction strength  $g_{1D}=-2\hbar^2/(ma_{1D})$ can be tuned from the weakly interacting regime ($g_{1D}\rightarrow \pm0$) to the strong coupling regime ($g_{1D}\rightarrow \pm \infty $) via Feshbach
resonances or confinement-induced resonances~\cite{Olshanii}. $g_{1D}>0$ ($<0$) represents repulsive (attractive) interaction. As usual, we define the dimensionless  interaction parameter $\gamma =c/n$ where $c=mg_{1D}/\hbar ^{2}=-2/a_{1D}$ and $n$ is the density. For our convenience in numerical calculation, we set $2m=\hbar =1$.  Full thermodynamics is accessible through the TBA equations derived by Takahashi~\cite{Tak70} and by others \cite{Schlottmann:1993,Hubbard-book,GuaBLB:2007} using the Yang-Yang method~\cite{Yang-Yang}. The solutions to the TBA equations have been studied for a variety of physical properties~\cite{GuaBLB:2007,YCLRG:2015,Guan-Ho,Patu:2015}. The experimental realizations of 1D Fermi gas~\cite{Pagano,Liao:2010,Wenz:2013,Murmann:2015} further provide an ideal testing ground for our understandings of few- and many-body physics~\cite{GuaBL13}.

{\em Equation of State (EOS).} We first derive the universal behaviour of attractive and repulsive 1D Fermi gases at high temperatures. Solving the TBA equations by proper iterations~\cite{Supp}, we find that the pressure $p$ and the density $n$ can be written as (valid for both attractive and repulsive)
\begin{align}
p \lambda_T^3 = &~8\pi \cosh ({H}/{2T}) z + \{2\sqrt{2}\pi \exp(\lambda^2)\left[1-{\rm erf}(\lambda) \right]\nonumber\\
&~- 4\sqrt{2} \pi \cosh^2 ({H}/{2T})\} z^2+O ( z^3), \label{pr}\\
n \lambda_T = &~2 \cosh({H}/{2T})z +\{\sqrt{2}\exp({\lambda^2})\left[1-{\rm erf}(\lambda) \right]\nonumber\\
&~- 2\sqrt{2} \cosh^2({H}/{2T})\} z^2+ O(z^3), \label{n}
\end{align}
where $z=\exp(\mu/T)$ is the fugacity and $\lambda_T=\sqrt{h^2/(2\pi mk_{\rm B} T)} $ is the thermal de Broglie wavelength. $\lambda=\text{sign}(c) \sqrt{c^2/(2T)}$ and ${\rm erf}$ denotes the standard error function. For the  balance case when $H=0$,  the EOS was computed in ref.~\cite{Hoffman:2015} and agrees with our results. For explicit comparison~\cite{Supp}, we note that different units and conventions were used in ref.~\cite{Hoffman:2015}. Here we derive for the first time the high temperature EOS for the 1D Fermi gas with an arbitrary polarization for both repulsive and attractive interactions.   

\begin{figure}
\begin{center}
\includegraphics[width=\columnwidth]{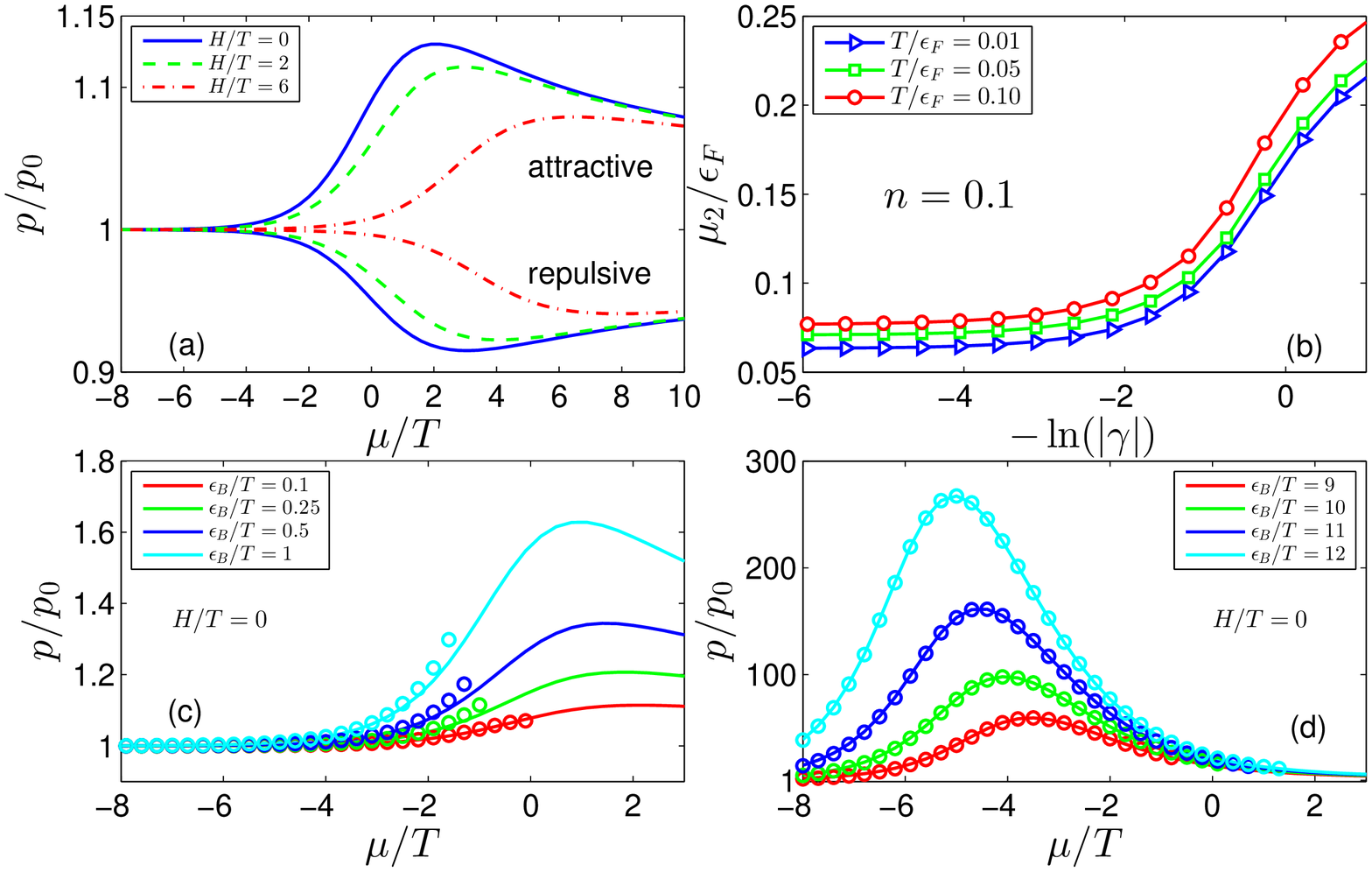}
\end{center}
\caption{(Color online) The dimensionless pressure ratio of  $p/p_0$  vs. $\mu / T$ for an 1D Fermi gas. Here $p_0$ is the pressure for the free fermions. 
   (a)  shows the numerical TBA  result for the EOS of the polarized Fermi gas with repulsive (lower branch) and attractive (upper branch) interactions for different magnetic fields~\cite{Supp}. 
   (b) The rescaled effective chemical potential of bound pairs $\mu_2/\epsilon_F$ as a function of  $\ln |\gamma|$ for $H=0$.
(c) At high temperatures/weak coupling,  the curves for this ratio show agreement between the analytical result  Eq. (\ref{pr}) (circles) and the numerical  TBA result (solid lines) for $H=0$. 
(d)  The ratio  $p/p_0$  vs $\mu / T$ for  the strongly attractive Fermi gas with different interacting strengths:  solid lines stand for  the numerical  TBA result while the circles denote the analytical result  from Eq. (\ref{EOS-polylog}). 
}
\label{fig:EOS-P}
\end{figure}

In Fig.(\ref{fig:EOS-P}), we show the normalized pressure $p/p_0$ as a function of $\mu/T$ for both repulsive and attractive gases. Here $p_0$ is the pressure of the corresponding non-interacting gas. In Fig. \ref{fig:EOS-P} (a), we note the opposite non-monotonicities of $p/p_0$ for the attractive and repulsive  gases, which are due entirely to interaction effects, where we choose $c=1$ for our numerical calculations. The deviation is smaller for larger magnetic fields, since the polarization of the gas is larger, leading to smaller interaction effects. In Fig.(\ref{fig:EOS-P}) (c,d), we show the perfect agreement between Eq.(\ref{pr}) and the numerical solution of TBA equations for various interaction strength at zero field $H=0$ at high/low temperatures.

The situation changes dramatically at low temperatures, in particular, for the attractive gas where  the formation of Cooper pairs induces fundamental changes in the thermodynamic behaviour of the system. In this regime, it is convenient to define the effective chemical potentials for the unpaired fermions $\mu_{1}= \mu+H/2$ and for the Cooper pairs $\mu_{2}= \mu +\epsilon_{\rm B}/2$, where the binding energy $\epsilon_{\rm B}=2a_{\rm 1D}^{-2}$ can be used to characterize the crossover from the strong to weak pairing regimes. In particular, we find for $ \epsilon_{\rm B}\gg T$, the pressure can be written as a sum of two components: $p=p_1+p_2$, where $p_1$ is the pressure for unpaired fermions and $p_2$ for pairs, given explicitly in~\cite{Guan-Ho} 
\begin{equation}
 p_{1}= F^{1}_{\frac{3}{2}}\left[1
+\frac{p_2 }{4|c|^3} \right], \,p_{2}= F^{2}_{\frac{3}{2}}\left[1+\frac{4p_1}{|c|^3}
+\frac{p_2 }{4|c|^3}\right], \label{EOS-polylog}
\end{equation}
respectively. Here $F^{r}_{a}=-\sqrt{{r}/{4\pi}}T^a{\rm Li}_a[-\exp({A_{r}/T})]$ ($r=1,2$) with
$A_{1} = \mu_1-\frac{2p_2}{|c|} +
\frac{1}{4|c|^3}F^{2}_{\frac{5}{2}} +Te^{-\frac{H}{T}}e^{-\frac{J}{T}}I_0\left(\frac{J}{T}\right)$
and  $A_{2} = 2\mu_2-\frac{4p_1}{|c|}-\frac{p_2}{|c|}+
\frac{8}{|c|^3}F^{1}_{\frac{5}{2}}+
\frac{1}{4|c|^3}F^{2}_{\frac{5}{2}}$.
In the above equations $\mathrm{Li}_{n}(x)=\sum_{k=1}^\infty{x^k}/{k^n}$ is the polylogarithm function, $J=2p_{1}/|c|$ and $I_0(x)=\sum_{k=0}^{\infty}\left(x/2\right)^{2k}/{(k!)^2}$. For more discussions, see ref.~\cite{Supp}. For interaction strength $\epsilon_{\rm B} \sim  \epsilon_F$, where $\epsilon_F=n^2\pi^2$ is the Fermi energy,  the effective chemical potential  $\mu_2$ increases quickly, approaching $\epsilon_F$, as shown in Fig.\ref{fig:EOS-P} (b). This feature was also observed in the recent experiment on the 2D attractive Fermi gas~\cite{Boettcher:2015}.  A good  agreement  between the analytical and numerical results  in high (Fig.~\ref{fig:EOS-P}(c))  and  low  (Fig.~\ref{fig:EOS-P}(d)) temperature regimes shows that the strongly attractive Fermi gases with polarization can be regarded as the mixture of two ``free" gases of bound pairs and single atoms as long as $\epsilon_{\rm B} \gg T$. In fact, the effects from the pair-pair and pair-unpaired fermion interactions are already encoded in the effective chemical potentials. We note that in the strong pairing regime, the effective chemical potential $\mu_2$ approaches a constant, which indicates that pairs are tightly bound.
%
 


\begin{figure}[h]
\begin{center}
\includegraphics[width=\columnwidth]{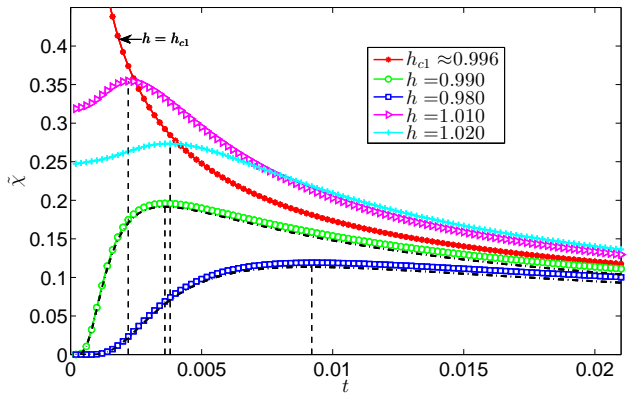}
\end{center} 
\caption{Dimensionless susceptibility vs. temperature for the attractive gas with  the chemical potential $\mu=-0.249$ and  the coupling strength $c=-1$. For external field $h<h_{c1}=0.996$, the susceptibility of Eq.\ref{sus-gapped} (thick black-dashed lines) is in  excellent agreement with numerical result  from the TBA equations \cite{Supp}.  The red dotted line shows the onset of susceptibility  at $h=h_{c1}$. For $h>h_{c1}$ the gap vanishes with an antiferromagnetic behaviour. 
 }
\label{fig:sus}
\end{figure}

{\em Characterizing quantum phases in an attractive gas when $H\neq 0$}. For a fixed value of chemical potential, the phase diagram of an attractive Fermi gas consists of three phases:  a fully-paired phase P for $H<H_{c1}$,  a fully-polarized ferromagnetic phase F for $H>H_{c2}$, and a partially polarized phase PP  for the intermediate field $H_{c1}<H<H_{c2}$~\cite{Orso:2007,Hu:2007,GuaBLB:2007}. Here $H_{c1,c2}$ are the lower and upper critical fields. To characterize these phases and phase transitions between them, we calculate the following two quantities: (1) the magnetic susceptibility $\chi=\partial M/\partial H$ that characterizes the magnetic properties of the system and (2) the ``contact susceptibility", $C_{h}\equiv\partial C/\partial H$ that characterizes how the singlet pair formation evolves with external magnetic field. Here $C$ is the contact of the 1D attractive Fermi gas, measuring the probability of short-range singlet pairs in the system \cite{Barth:2011}. It is important to note that while $\chi$ is related to the one-body density matrix of the system, $C_{h}$ is determined by the two-particle density matrix and provides a unique quantity to characterize the pair fluctuation of the system. The contact $C$ also  manifests the criticality of the gas near the critical point~\cite{CJGZ}. In fact, as we will show later, there exists a unique link between $C$ and $M$.

{\em (1) Magnetic susceptibility}. The 1D Fermi gas offers an ideal platform for the exploration of quantum magnetism~\cite{Pagano,Liao:2010,Wenz:2013,Murmann:2015}. For systems with an excitation gap $\Delta$, the magnetic susceptibility $\chi$ usually takes the form $\chi\sim(1/\sqrt{T})\exp(-\Delta/T)$ \cite{Johnston:2000}, indicative of a dilute set of magnons at low temperatures. Here we prove explicitly that the TBA equations for an attractive Fermi gas determine precisely such an elegant behaviour for $H<H_{c1}$.

Defining the dimensionless quantities $\widetilde{\mu}=\mu/\epsilon_{\rm B}, \,h=H/\epsilon_{\rm B}$, $t=T/\epsilon_{\rm B}$ and $\widetilde{\Delta}=\Delta/\epsilon_{\rm B}$, the dimensionless susceptibility $\widetilde{\chi}=\epsilon_{\rm B} \chi/|c|$ for the gapped Fermi gas can be written as
\begin{equation}
\widetilde{\chi}=-\frac{1}{4\sqrt{2}}f_{-\frac{1}{2}}\left(\frac{\widetilde{\Delta}}{t}\right)\left[ 1+3f_{\frac{1}{2}}\left(\frac{\widetilde{\Delta}}{t}\right)+3 f_{\frac{1}{2}}\left(-\frac{\widetilde{A}_2}{t}\right)\right],
\label{sus-gapped}
\end{equation}
where $f_n(x)={t^n}/{(2\sqrt{\pi})}\Li_{n}(-\exp(-x))$ and $\widetilde{A}_2\equiv A_2/\epsilon_{\rm B}$. At low temperatures when $t\ll \widetilde{\Delta}$, $\widetilde{\chi}$ reduces to a surprisingly simple expression
\begin{equation}
\widetilde{\chi}\approx  \frac{1}{8\sqrt{2\pi}\sqrt{t}}e^{-\widetilde{\Delta}/t}. 
\label{sus-formula}
\end{equation}
The dimensionless energy gap $\widetilde{\Delta}\equiv\Delta/\epsilon_{\rm B}$ for a strongly attractive Fermi gas can be written as (up to the order $1/c^3$) 
  \begin{equation}
\widetilde{\Delta}=-\tilde{\mu}_1+\frac{16{\tilde{\mu}_2}^{\frac{3}{2}}}{3\sqrt{2}\pi}-\frac{16{\tilde{\mu}_2}^{2}}{3\pi^2} +\frac{112{\tilde{\mu}_2 }^{\frac{5}{2}}}{9\sqrt{2}\pi^3 }-\frac{32{\tilde{\mu}_2}^{\frac{5}{2}}}{15\sqrt{2}\pi}
\label{gap}
\end{equation}
with $\tilde{\mu}_1=\tilde{\mu}+h/2$ and $\tilde{\mu}_2=\tilde{\mu}+1/2$. 

Eq.(\ref{sus-formula}) presents a general signature for a spin gapped systems in 1D systems~\cite{Dagotto:1999,Bat2007}, and we derive the explicit form for the first time from Bethe Ansatz.  This relation shows that the susceptibility decays exponentially with decreasing temperature with a pre-factor proportional to $1/\sqrt{T}$. The temperature dependent susceptibility is presented in Fig.~\ref{fig:sus}, where an excellent agreement between the numerical result and the analytical formula Eq.(\ref{sus-formula}) is observed for the external field $h<h_{c1}=0.996$. The typical rounded peaks show a characteristic of spin gapped phase for $h<h_{c1}$, where the gap given by Eq.(\ref{gap}), agrees fully with numerical result in Fig.~\ref{fig:sus}. As expected, the suscetibility diverges at $h=h_{c1}$ where magnetic phase transition occurs \cite{Vekua:2009}. However, for $h>h_{c1}$, the system turns into a partially polarized phase with the susceptibility satisfying an additivity rule at low temperature, i.e. $\chi^{-1} =\chi_1^{-1} +\chi^{-1}_2$, where $\chi_{r} =r(\mu_{\rm B} \mathfrak{g} )^{2} \left({\partial n_{r}}/{\partial \mu_{r}}\right)_{n,c}$ with $r=1,2$.  This additivity rule is a characteristic of the TLL, which was found in~\cite{YCLRG:2015,Guan:2013PRL}. Here $\mu_B$ is the Bohr magneton and $ \mathfrak{g}$ is the Land\'{e} factor. 

\begin{figure}
\begin{center}
\includegraphics[width=\columnwidth]{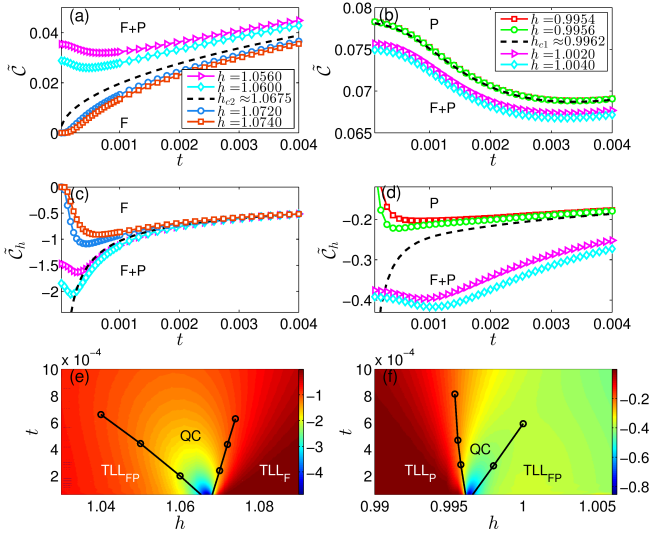}
\end{center} 
\caption{The dimensionless contact $\widetilde{C}$ and contact susceptibility $\widetilde{C}_h$ as a function of dimensionless temperature $t$ for the attractive Fermi gas with $\tilde{\mu} = -0.498$. (a) and (c):   $\widetilde{C}$ and $\widetilde{C}_h$ for different magnetic fields near  the upper critical field $h_{c2}=1.0675$. (b) and (d):  $\widetilde{C}$ and $\widetilde{C}_h$ for different magnetic fields near the lower critical field $h_{c1}=0.9962$. (e) and (f)  show the contour plots of the contact susceptibility as a function of the temperature at the upper and the  lower critical points, respectively. The black lines show the minimums of  the contact susceptibility . }
\label{fig:contactT}
\end{figure}

{\em (2) Contact susceptibility}. Physically, the contact characterizes the probability of finding two particles entangled in a spin-singlet state (pair) at short distances.  It establishes universal relations among many physical quantities ~\cite{Tan:2008,Braaten:2008,Zhang:2009,Barth:2011,Sagi:2012,Bar:2015}, such as the tail of momentum distribution, the interaction energy, the pressure and the dynamic structure factor at high frequencies~\cite{Braaten:2008}. Very recently it was also shown that the contact exhibits the universal scalings at the second order phase transitions~\cite{CJGZ}. As a function of temperature, the evolution of $C$ indicates how the singlet pair formation is affected by thermal fluctuations. Furthermore, the ``contact susceptibility", $C_h\equiv\partial C/\partial H$ measures the likelihood of pair breaking by the external magnetic field \cite{Thekkadath:2015} and, as we have emphasized before, is related to the two-body density matrix, in contrast to the magnetic susceptibility.

In Fig.\ref{fig:contactT}, we show the dimensionless contact $\widetilde{C}\equiv C/\epsilon_{\rm B}^2$ and $\widetilde{C}_h\equiv C_h/\epsilon_{\rm B}$ as a function of the dimensionless temperature $t\equiv T/\epsilon_{\rm B}$. There are several features to be noted. (1) In both the fully-paired and partially polarized phase, the contact exhibits non-monotonic dependences on temperature. As one increases temperature from zero, the contact starts with a finite value and initially decreases due to thermal fluctuations, but finally increases due to decreasing polarization. On the other hand, for $H>H_{c2}$, in the fully polarized phase, the contact starts from zero and increases monotonically with temperature. (2) The contact susceptibility, on the other hand,  starts from zero for both the paired and full polarized phase, reflecting the gapped nature of the spin excitations. The temperature at which the contact susceptibility reaches maximum indicates the strongest pair fluctuations in the system. The minimums  of the $\widetilde{C}_h$ also mark the characteristic  temperatures which distinguish the TLL phases from the quantum critical regime.

{\em Bounds for contact in a magnetized gas.} The fact that both magnetic and contact susceptibilities can be used to characterize the critical behaviour of the system and give rise to the same quantum critical regime implies that there is a fundamental relation between these two quantities. To see that, let us first consider the high temperature limit. In this case, the contact is proportional to the probability of the random collisions between two fermions with different spins, i.e. $C \propto (N_{\downarrow}/N)(N_{\uparrow}/N)=(1-P^2)/4$ where $P\equiv (N_{\uparrow}-N_{\downarrow})/N=2M/N$ is the polarization. This represents the limit of least (spin singlet) paired Fermi gas. On the other hand, in the strong coupling limit where fermions form tightly bound molecules, the contact should be proportional to the number of molecules $N_{\downarrow}$, and as a result $C\propto 1-P$. This represents the limit of a strongly (spin singlet) paired states. In other words, if we define the contact for $P=0$ as $C_{\rm max}$, then $C/C_{\rm max}$ should lie between two curves $1-P$ and $1-P^2$. In three-dimension, this relation is conjectured to hold and is shown to be so with a large-$N$ calculation~\cite{Bardon:2014}. For 1D Fermi gas, based on the TBA equation, we prove that this is indeed the case.

In the grand canonical ensemble, the Tan's contact can be calculated as $C=-(c^2/2) {\partial p}/{\partial c}$. By analytically solving the TBA equations for various cases,  the contact can be written as (for both repulsive and attractive cases)
\begin{widetext}
\begin{eqnarray}
 \mathcal{C}=\left\{\begin{array}{ll}
  \frac{2n^4\pi^2}{3}(1-P)[1-\frac{6n}{c}(1-P)+\frac{24n^2(1-P)^2}{c^2}-\frac{12n^2\pi^2}{5c^2}] , & c\gg1, \,\,P>0.5,\,\,T=0\\
      \frac{2\pi^2n^4 \ln2}{3} \left[1-\frac{6n\ln{2}}{c}+\frac{24(\ln{2})^2 n^2}{c^2}-\frac{6\pi^2\zeta(3)n^2}{5c^2\ln{2}} \right]-\frac{c^2}{8\pi^2n^2}T^2,& c\gg 1,\,\, P=0, \,\, T\ll \mu\\
   \frac{1-P^2}{4}c^2n^2,& |c|\ll 1,\,\,T=0\\
     \frac{n|c|^3}{4}  ( 1 - P) \left\{ 1 + \frac{\pi^2n^3}{|
  c |^3} \left[ \frac{1}{24}  ( 1 - P)^2  ( 1 + 3 P) 
  + \frac{8}{3} P^3 \right] \right\}+\frac{4P^2-P+1}{6(1-P)P}T^2,&c\ll -1 \,\,P\ne 0,1\,\,T\ll \mu,\\
\frac{c^2 n^2 ( 1 - P^2)}{4}  \left\{ 1 -
  \frac{\sqrt{\pi}c}{\sqrt{2T}} e^{\frac{c^2}{2 T}}\left[1-{\rm erf}( \frac{c}{\sqrt{2T}}) \right]
  \right\}, & T\gg c^2.
 \end{array}
\right.\label{contact-pp}
\end{eqnarray}
\end{widetext}

\begin{figure}[tb]
\begin{center}
\includegraphics[width=\columnwidth]{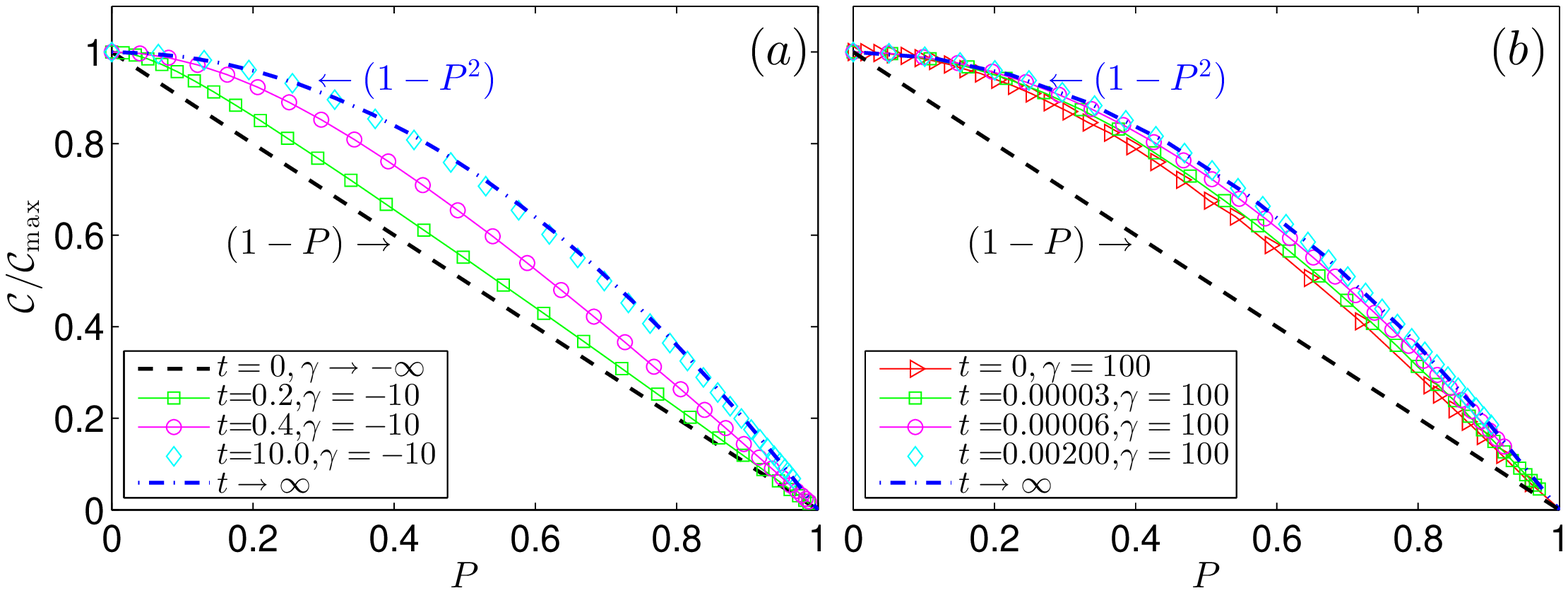}
\end{center} 
  \caption{The contact ratio $C/C_{\rm max}$ vs polarization $P$. Here $C_{\rm max}$ is the corresponding maximum value of the contact at $P=0$. 
   Left panel: for an attractive interaction with  $\gamma =-10$. Right panel: for a repulsion with  $\gamma=100$.  
   The ratio in the two interacting regimes are boundbetween the two limiting cases, {\it viz}. $1-P^2$ and $1-P$.}
  \label{fig:contact-p}
\end{figure}

In Fig.\ref{fig:contact-p}, we show $C/C_{\rm max}$ for various values of temperatures in the attractive case (left pannel) and the repulsive case (right pannel). They all lie within the asymptotic curves $1-P$ and $1-P^2$. As one increases the temperature, the curve moves gradually towards $1-P^2$, as we have argued intuitively above. The relation between $C$ and $M$ (or $P$) also provides a unique angle to understand the correlation  of Fermi gas in higher dimensions, namely the extent to which fermions are  entangled in a singlet pair in the system~\cite{Thekkadath:2015}.

In summary, we have characterized the quantum states of a polarized 1D Fermi gas by obtaining the exact EOS, magnetic susceptibility, the contact and contact susceptibility for both repulsive and attractive interactions. We have shown that quantitative measurement of pair fluctuation through contact susceptibility also marks the quantum critical regime and have derived exact bounds for the contact in a spin polarized system. These benchmarking thermodynamics for the 1D crossover from weak to strong pairing regimes [Fig.~\ref{fig:EOS-P}], susceptibility [Fig.~\ref{fig:sus}], and the upper and lower bounds for the contact [Fig.~\ref{fig:contact-p}], are common for both 2D/3D and 1D systems. These properties  can now be probed in quasi-1D experiments~\cite{Pagano,Liao:2010,Murmann:2015} and could also shed new light into the universal behaviour of interacting fermions in higher dimensions~\cite{Bardon:2014,Boettcher:2015,Fenech:2015}. 

\noindent

{\em Acknowledgments.} We thank T.-L. Ho  for helpful discussions. This work is supported by the NNSFC under grant numbers 11374331 and the key NNSFC grant No. 11534014 and by the National Basic Research Program of China under Grant No. 2012CB922101. SZ is supported by Hong Kong Research Grants Council (General Research Fund, HKU 17306414, Collaborative Research Fund, HKUST3/CRF/13G) and the Croucher Innovation Awards.

\end{document}